\documentclass[conference]{IEEEtran}
\makeatletter
\def\ps@headings{%
\def\@oddhead{\mbox{}\scriptsize\rightmark \hfil \thepage}%
\def\@evenhead{\scriptsize\thepage \hfil \leftmark\mbox{}}%
\def\@oddfoot{}%
\def\@evenfoot{}}
\makeatother
\usepackage{xcolor}
\usepackage{times}
\usepackage{psfrag}
\usepackage{epsfig}
\usepackage{algorithm}
\usepackage{algorithmic}
\usepackage{multicol}
\usepackage{multirow}
\usepackage{setspace}
\usepackage{cite}
\usepackage{graphicx}
\usepackage{array}
\usepackage{afterpage}
\usepackage{amssymb}
\usepackage[cmex10]{amsmath}
\usepackage{mdwmath}
\usepackage{mdwtab}
\usepackage[tight,footnotesize]{subfigure}
\usepackage{mdwlist}
\usepackage{tikz}

\usepackage{geometry}
\usepackage{yhmath}
\usepackage[amssymb]{SIunits}
\begin{document}
\bibliographystyle{IEEEtran}
\title{Physical Layer Key Generation in 5G Wireless Networks}
\author{\IEEEauthorblockN{Long Jiao, Ning Wang, Pu Wang, Amir Alipour-Fanid, Jie Tang, Kai Zeng}
\IEEEauthorblockA{Department of Electrical and Computer Engineering\\George Mason University, Fairfax, VA, US 22030\\
E-mail: \{ljiao, nwang5, pwang20,aalipour, jtang20, kzeng2\}@gmu.edu}}
\maketitle
\begin{abstract}
The bloom of the fifth generation (5G) communication and beyond serves as a catalyst for physical layer key generation techniques.
In 5G communications systems, many challenges in traditional physical layer key generation schemes, such as co-located eavesdroppers, the high bit disagreement ratio,  and high temporal correlation, could be overcome.
This paper lists the key-enabler techniques in 5G wireless networks, which offer opportunities to address existing issues in physical layer key generation.
We survey the existing key generation methods and introduce possible solutions for the existing issues.
The new solutions include applying the high signal directionality in beamforming to resist co-located eavesdroppers, utilizing the sparsity of millimeter wave (mmWave) channel to achieve a low bit disagreement ratio under low signal-to-noise-ratio (SNR),  and exploiting hybrid precoding to reduce the temporal correlation among measured samples.
Finally, the future trends of physical layer key generation in 5G and beyond communications are discussed.
\end{abstract}
\begin{tikzpicture}[overlay, remember picture]
\path (current page.north east) ++(-5,-0.5) node[below left] { This work has been accepted by the IEEE Wireless Communications
Magazine. };
\end{tikzpicture}

\section{Introduction}



Over the past few years, the fifth generation (5G) communication has come a long way, thanks to the booming wireless technologies, such as millimeter wave (mmWave), massive MIMO, highly directional beamforming, and hybrid precoding \cite{yang2015safeguarding1}.
Compared with the current 4G network, 5G networks can satisfy the increasing demand by providing a high data rate, ultra reliable low latency, and massive machine type communications.
In the design of 5G networks, providing reliable and secure communication service is one of the top priorities.
Specifically, in this article, we focus on one of the promising physical layer security mechanisms - physical layer key generation, and explore the benefits offered by the new technologies applied in 5G wireless networks.

Different from the traditional Diffie-Hellman (D-H) key exchange mechanism, physical layer key generation mechanisms do not require expensive computation and have the potential to achieve information-theoretic security \cite{maurer1993secret}.
That is, the secrecy of the generated key is not dependent on the hardness of a computational problem but relies on the physical laws of the wireless fading channels. 
For instance, wireless devices measure highly correlated wireless channel characteristics and use them as shared random sources to generate a shared key \cite{JiDu2017TVT}.
In theory, in a rich multi-path scattering environment, a passive eavesdropper who is more than a half-wavelength away from legitimate users will obtain un-correlated channel measurements, thus cannot infer much information about the generated key.

Physical layer key generation has gained much attention in literature \cite{zeng2015physical}. Based on the number of antennas in the transceivers, physical layer key generation schemes under sub-6GHz can be generally classified into two categories: single antenna and MIMO based key generation.
In single-antenna based mechanisms, various channel characteristics have been proposed to generate the secret key, including received signal strength (RSS) \cite{zeng2010exploiting}, channel state information (CSI) \cite{zeng2015physical} and angle of arrival (AoA)\cite{jiao2018physical}.
In MIMO based mechanisms, works in \cite{MIMOK} conducted an indoor MIMO measurement in the 2.51-2.59GHz band and studied the number of available key bits in both line-of-sight (LoS) and non-line-of-sight (NLoS) environments. 
In \cite{MIMOK}, a theoretical upper bound for the maximum size of the generated secret key is derived based on the mutual information of channel estimates at the two legitimate nodes.
But the number of antennas considered is still relatively small and the carrier frequency is not as high as mmWave.
Furthermore, all the theoretical analysis are based on the Gaussian channel assumption which cannot be directly utilized in mmWave channels because of their unique scattering nature \cite{rappaport, zhang20}.

Although significant efforts and progress of physical layer key generation have been made in recent years, many issues or challenges still remain elusive.
For example, most existing key generation schemes cannot combat the co-located eavesdropping in sub-6GHz systems.
That is, a common assumption that eavesdroppers locate at least a half of wavelength apart from legitimate users is required in most key generation schemes.
Furthermore, some existing works have a high bit disagreement ratio in the low SNR regime, which leads to a high reconciliation overhead and low efficiency of key generation.
In addition, a high probing rate is chosen to increase the key generation rate in most of the existing works, however, it will lead to the high temporal correlation among samples along with a large number of repeated quantization bits.

With unique features, 5G and beyond technologies may offer a better solution to such aforementioned issues or challenges remained in existing works. 
For instance, the high directionality enabled by massive MIMO based beamforming is exploited in \cite{jiao2018secretbeam} to defend against co-located eavesdroppers in key generation.
Besides, authors developed a scheme by utilizing the sparsity of mmWave channel to estimate virtual angle of arrival (AoA) and angle of departure (AoD) \cite{jiao2018physical}.
Due to the sparsity of mmWave channel, the scheme is anti-noise and can achieve a low bit disagreement ratio. 
Moreover, based on the fact that hybrid precoding can form mmWave beams with multiple resolutions, it can be adopted in the channel probing stage to reduce the temporal correlation among samples.


By identifying possible solutions and benefits offered by 5G communication technologies, the purpose of this paper is to provide new insights about the physical key generation in 5G wireless networks and is expected to advance and stimulate the corresponding research under the context of 5G and beyond communication systems.
It should be noted that existing survey and tutorial papers \cite{zeng2015physical} on physical layer key generation mainly focus on sub-6GHz systems. 
A comprehensive study of physical layer key generation in 5G wireless networks, identifying challenges and opportunities, is still lacked, which is the major contribution of this paper. 	

In Section \ref{Tech}, we introduce the basic of key generation and analyze limitations of existing works.
In Section \ref{tech-5g}, the specific benefits or improvements accompanied with 5G communication systems are discussed.
In Section \ref{Improving}, we show the benefits of 5G communication technologies on physical layer key generation with three typical cases.
The future trends and research topics are discussed in Section \ref{Future}, while the conclusions are given in Section \ref{Conclusion}.

\section{Key Generation and Existing Issues}
\label{Tech}

In this section, we review the typical key generation process and analyze the limitations of existing schemes. Here, two devices, denoted by Alice and Bob, wish to generate a common secret key based on channel measurements.
To extract the secret key, Alice and Bob generally perform five steps: channel probing, randomness extraction, quantization, information reconciliation, and privacy amplification.

\subsection{Typical Key Generation Process}
\begin{itemize}
\item {\textbf{Channel Probing:}}
In this step, Alice and Bob exchange channel probing signals with each other to collect enough channel measurements as a shared random source.
The channel measurements can be CSI, phase, or AoA and AoD.
If the channel reciprocity holds, the received measurements at Alice and Bob are highly correlated.

\item {\textbf{Randomness Extraction:}}
The received signals at Alice and Bob may contain deterministic parts that can be inferred by an attacker.
Randomness extraction is adopted here to remove deterministic parts and extract randomness of the channel fading.

\item {\textbf{Quantization:}}
This is used to quantize the random channel measurements into the binary bits.

\item {\textbf{Information Reconciliation:}}
Information reconciliation is a form of error correction carried out between Alice and Bob to ensure identical keys generated separately on both sides.
The extracted bits at Alice and Bob sides after quantization are usually not identical due to imperfect reciprocity and measurement noise.
During reconciliation, Alice and Bob exchange side information to correct errors, and a certain amount of bit information could be revealed to the eavesdropper.

\item {\textbf{Privacy Amplification:}}
This step is used to eliminate the leaked bits during channel probing and reconciliation.
After this step, eavesdropper's partial information by observation will be reduced largely.
\end{itemize}

\subsection{Limitations in Existing Schemes}

Although there have already been various studies of physical layer key generation for sub-6GHz systems, there remain several grave challenges and limitations for the existing key generation schemes in practice. We list three challenges existed in previous works.

\begin{itemize}
	\item \textbf{Co-located Attacks:}
The correlation of channel measurements is determined by the distance between the eavesdropper and Alice/Bob.
In the existing works\cite{zeng2015physical,JiDu2017TVT}, if the eavesdropper is co-located to Alice or Bob, he/she will observe channel measurements that are highly correlated with Alice or Bob.
Therefore, most existing works are vulnerable to co-located eavesdroppers.
How to design a security-proven physical key generation scheme under co-located attacks has not been well explored.
\item \textbf{High Bit Disagreement Ratio in Low SNR Regimes:}
The similarity of channel measurements at Alice and Bob is highly determined by the signal-to-noise-ratio (SNR) level.
In low SNR regimes, the noise leads to a high bit disagreement ratio after quantization, which could cause a high reconciliation overhead and thus lead to a low key generation rate.
Generally, existing works try to decrease the bit disagreement ratio by enhancing the SNR or reducing the quantization level.
Obviously, a scheme achieving a low bit disagreement ratio in low SNR regimes is desired.
\item \textbf{High Temporal Correlation:}
In previous works, in order to increase the key generation rate, a high probing rate is adopted in channel probing stage to obtain more channel measurements, which can be quantized into more binary bits. However, within the coherence time, the temporal channel measurements are highly correlated, which leads to duplicated key bits.
Apparently, a scheme that can maintain a high probing rate while reducing the temporal correlation is of great interest.
\end{itemize}

%
%
\begin{figure}
	\centering
	\includegraphics[width=0.47\textwidth,height=0.39\textwidth]{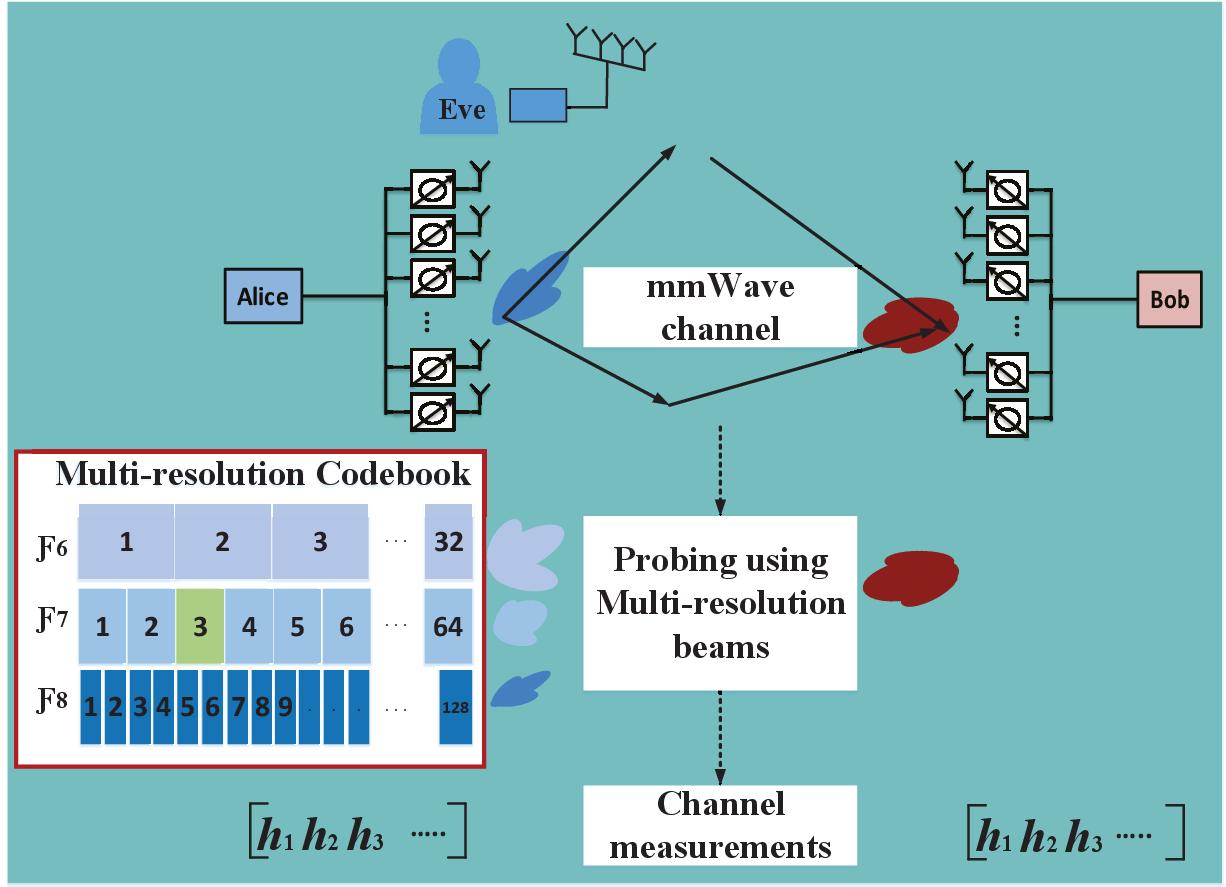}	
  \caption{Key generation for mmWave Massive MIMO with multi-resolution beams}
	\label{Hybridprecoding}
\end{figure}

\section{5G-Enabled Secure and Efficient Key Generation}
\label{tech-5g}
Before we point out the solution to the existing issues in key generation, we first introduce some disruptive technologies in 5G wireless networks. The opportunities and challenges of key generation based on these technologies are then discussed.

\subsection{mmWave Communication}
In 5G wireless networks, mmWave communication systems with the frequency range of 30-300 GHz, are considered as a promising solution to increase communication capacity, i.e., mmWave cellular systems will enable gigabit-per-second data rates thanks to the large bandwidth available at mmWave frequencies.
But Radio Frequency (RF) signals in mmWave communication has a short transmission range due to its heavy free-space path loss.

That may restrict the ability of the eavesdroppers because passive eavesdroppers need to be aware of her/his distance to the legitimate user\cite{yang2015safeguarding1}.
In other words, the information leakage of the key generation process in mmWave communication is more sensitive to the location than the sub-6GHz systems.

Different from sub-6GHz systems, which model the correlation of channel as Jake's correlation model \cite{wei2013adaptive}, mmWave channel is location-specific and cannot be directly modeled as Jake's model. The proper eavesdropping model is under-explored.
Besides, the propagation of mmWave at high frequency has a higher outage probability compared with sub-6GHz systems.
The blockage effect of mmWave has effects on physical layer key generation and needs to be considered in the future analysis of secrecy key rate.

\subsection{Beamforming (BF) and Massive MIMO}
BF, typically realized with the antenna array, can be applied to provide a high antenna gain and mitigate the severe path loss at mmWave frequencies.
In BF, the phase of antenna elements is adaptively shifted to form a concentrated and directed beam pattern\cite{sun2014mimo}.

In sub-6GHz systems, the BF is usually performed with several antenna elements in an antenna array. In mmWave communication, the number of antenna elements in the antenna array is greatly increased to enlarge the antenna gain of BF. MIMO systems with a great number of antennas are named ''Massive MIMO''. Thanks to the small wavelength of mmWave signals, a large number of antenna elements can be packaged on a chip. For example, in \cite{zihir201560}, the author proposes a small antenna array ($40mm\times41mm$) with 256 antenna elements, which thus enables the tiny antenna array to be embedded in Tx and Rx.

Massive MIMO based BF brings many benefits for physical layer key generation. At first, the ability of eavesdroppers is limited. The power level in Massive MIMO is reduced which cuts the SNR received at eavesdroppers.
Apart from this, the massive MIMO can generate very narrow beams by focusing on legitimate users without spilling over the signal power in other directions \cite{wu2018survey}. Due to the narrow beamwidth, an eavesdropper without beam tracking capabilities has high probabilities to lose eavesdropping links due to the factors like slight rotation of the antenna array \cite{jiao2018secretbeam}. On the contrary, the beam tracking techniques are required at the eavesdroppers to maintain a high SNR while eavesdropping.

The main challenge accompanied with massive MIMO is the large channel estimation overhead.
In existing works, most of the schemes usually choose the CSI as common randomness for physical layer key generation, which, however, is nontrivial to obtain in mmWave systems with massive MIMO.
The large number of antenna elements causes large channel estimation overhead\cite{alkhateeb2014channel12}.
Consequently, if Alice and Bob need to do bi-directional probing for channel estimation, the efficiency of key generation is going to be affected by large channel estimation overhead, which is needed to be tackled with in future research.
New schemes to combat these issues need to be investigated.

\subsection{Hybrid precoding}
Due to the short wavelength of mmWave, massive MIMO is quite promising for mmWave systems since antenna elements can be packaged on a chip.
The increasing number of antenna elements requires a larger phase array shifter networks. However, the number of beamforming chains controlling phase shifter array cannot be increased linearly with the number of antenna elements.
In order to provide a high precoding gain with less RF chains, the hybrid precoding structure is proposed \cite{alkhateeb2014channel12}.

In this paper, we propose a physical layer key generation scheme by utilizing the hybrid precoding structure for mmWave Massive MIMO systems, as depicted in Fig. \ref{Hybridprecoding}. The hybrid precoder is the combination of an analog precoder in RF domain and a digital precoder in baseband domain. The RF precoder is composed of several analog phase shifter networks. In the digital domain, the baseband precoder is usually designed by using compressive sensing techniques. For each round of bi-directional probing, Alice/Bob would select the beams with different resolutions and angles. After all probings and getting enough samples, Alice and Bob would perform the remaining steps of physical layer key generation.

As depicted in Fig. \ref{Hybridprecoding}, a multi-resolution codebook is utilized for precoding in our proposed scheme and it enables Alice and Bob to provide the multi-resolution beams in the quantized angle space. Our scheme based on hybrid precoding can reduce the high temporal correlation under the high sampling rate in the following two aspects, the steering angle and the beam resolution.
At first, by extracting every multi-path in the sparse mmWave channels, hybrid-precoding based key generation scheme can reduce the high temporal correlation of the channel measurements. The multi-paths belonging to different clusters experience independent scattering effects and thus possess independent statistical information. Such effect thus can degrade the temporal correlation among channel measurements and improve the key generation rate. Secondly, by adopting different beam resolutions in each round of bi-directional probing, the temporal correlation can be further decreased. For each round of the bi-directional probing, the received signal in each round can be viewed as a weighted combination of multi-paths. By adjusting the beam resolution in each round, the weights on multi-paths can be affected. Selecting appropriate beam resolution can further reduce the temporal correlation. The performance of the proposed scheme will be given and discussed in section \ref{BeamResolution}.

Although hybrid precoding is promising in key generation, the challenges associated with hybrid precoding cannot be ignored.
The precoding gain of hybrid precoding is superposed by the analog precoder and baseband precoder. The precoding errors can be introduced due to the imperfection of analog components and thus need to be considered in the theoretical analysis of secrecy key rate.

\section{Improving Security of Physical Layer Key Generation in mmWave Massive MIMO Systems}
\label{Improving}

The implementation of physical layer key generation in mmWave massive MIMO communication systems is highly rewarding.
In this section, we will discuss the benefits from three specific cases: countering co-located eavesdroppers, achieving a low bit disagreement ratio under low SNR, and reducing the temporal correlation under the high probing rates.

\subsection{Countering Co-located Eavesdroppers with High Directional Beams}

In mmWave systems, transceivers deploy narrow beams with the high directionality to suppress the interference from neighbors, which means eavesdroppers on the sidelobes have a very low SNR.
In order to achieve a high SNR, the eavesdroppers in mmWave systems will approach legitimate transceivers as close as possible, which may bring threats for current physical layer key generation schemes, where it mainly relies on the location decorrelation to ensure security.

Enabled by the massive MIMO based beamforming techniques, the high directionality of beams offers a solution to counter the co-located eavesdroppers. In \cite{jiao2018secretbeam}, we developed a key generation scheme for mmWave massive MIMO communication systems called ''Secret Beam'', where small random perturbation angles on the beamformer are chosen as the common random source.
The overall process in ''Secret Beam'' includes the following steps: inter-perturbation, quantization, and XOR operation.

Due to the sparsity of mmWave channel and the narrow beam width of the massive MIMO communication system, the eavesdropper needs to approach to Alice or Bob so that the narrow beams from certain directions can be received. For the eavesdropper being co-located with Alice/Bob, Alice and Bob conduct the inter-probing strategy and adjust the perturbation on BF vectors to perform key generation. In the worst case, the eavesdropper may get the same bits with Alice or Bob but not both. However, the eavesdropper by no means can determine the key which is the result by XORing bits generated at Alice and Bob sides.


The mmWave massive MIMO system adopts analog phase-shifter with a single radio-frequency (RF) chain. We discussed the uniform planar array (UPA) adopted at Alice and Bob sides with two dimensions, $32 \times 16$ and $16\times 8$. The narrowband mmWave operates on 28 GHz and the channel reciprocity holds in the coherence time.
The mmWave channel contains a LoS path and an NLoS path, where the amplitude of NLoS path is typically 10dB weaker than the LoS path.

\begin{figure}
	\centering
	\includegraphics[width=0.48\textwidth,height=0.38\textwidth]{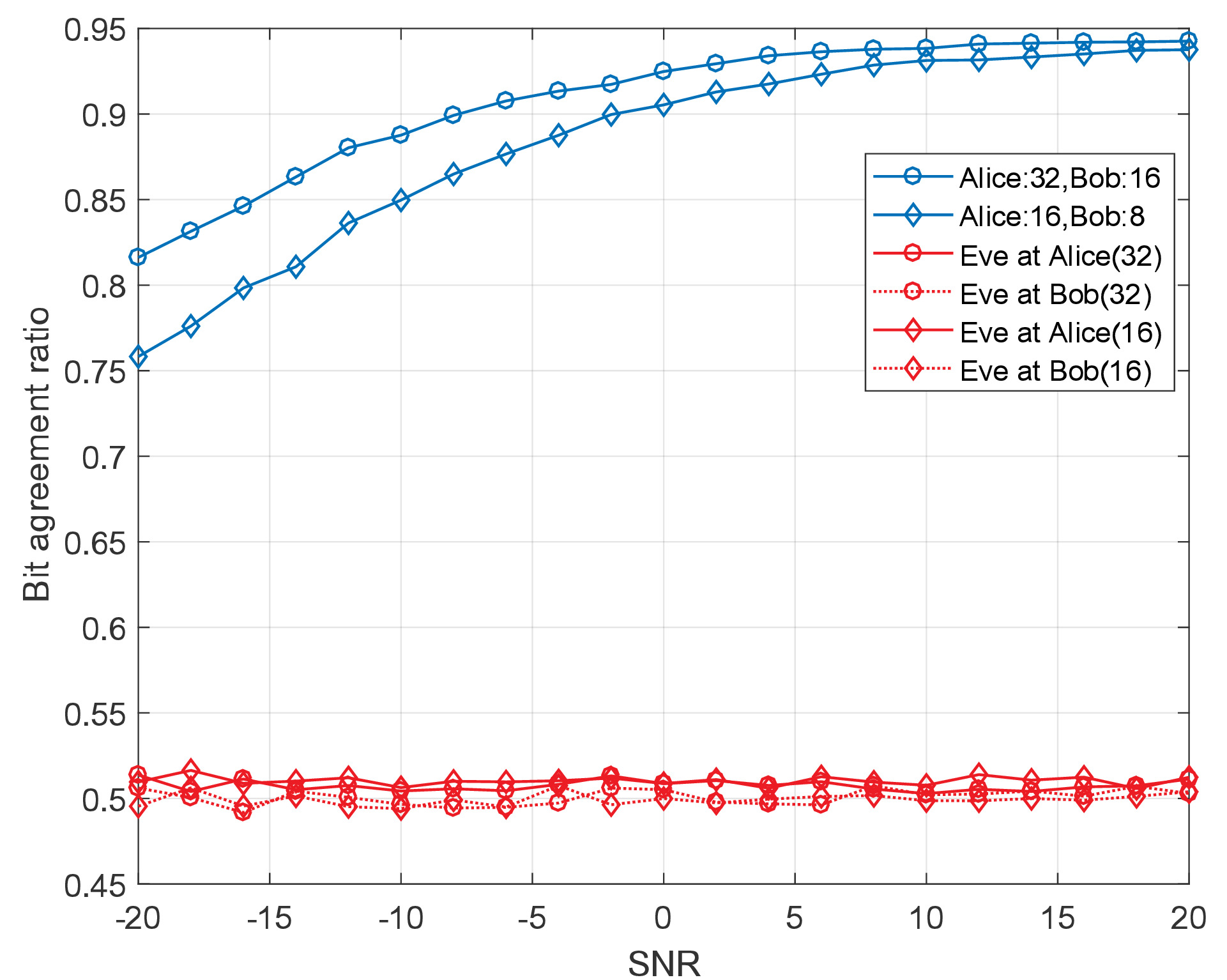}	
  \caption{The bit agreement ratio under the co-located eavesdroppers}
	\label{AttackersAgreementRatio}
\end{figure}

In Fig. \ref{AttackersAgreementRatio}, we compared the bit agreement ratio (BAR) in two cases with a fixed quantization level of 16. In the first case, the dimensions of UPA at Alice and Bob sides are 32 and 16, respectively (i.e., $32\times16$) and the blue curve marked with circles represents the BARs under different SNRs. In the second case, the blue curve marked with diamonds represents the BARs while adopting the smaller antenna number $16\times8$ at Alice and Bob side. We can observe that a higher BAR can be achieved with more antenna elements. There are four red curves represents four different cases, where the eavesdropper can either be co-located with Alice or Bob under two antenna dimensions ($32\times16$ or $16\times8$). Based on the four red curves, we can observe that the eavesdropper’s BAR in four cases after XOR operation is around $50\%$ under various SNRs. As a result, this scheme is robust against a co-located eavesdropper.

\subsection{Achieving a Low Bit Disagreement Ratio under Low SNR Regimes}

A high bit disagreement ratio after the channel sounding can greatly increase the reconciliation overhead and slow down the key generation rate.
For example, if the existing reconciliation cascade algorithm is used to reconcile two bit strings with a 10\% bit mismatch rate, 60\% of bits need to be discarded \cite{jiao2018physical}.
Therefore, having a low bit mismatch before reconciliation is critical in physical layer key generation.

In low SNR environments, existing works utilize a low quantization level to achieve the high key agreement ratio.
However, in these schemes, fewer bits are quantized due to the low quantization level, which results in a low key generation rate.
In contrast, we proposed a new key generation scheme in \cite{jiao2018physical} exploiting the sparsity of mmWave Massive MIMO channel and the virtual AoAs and AoDs to obtain a high bit agreement ratio (BAR) (99\%) even under low SNR regimes.
In the mmWave massive MIMO channel, virtual AoAs/AoDs are transformed from physical AoAs/AoDs after projecting the mmWave channel onto the unitary matrix, as the virtual channel matrix.
In this way, we can observe that the entries corresponding to physical AoAs/AoDs are with the large amplitudes, as well as virtual AoAs and AoDs associated with physical AoAs and AoDs.
On the other hand, if the entries in virtual channel matrix are not corresponding to physical AoAs/AoDs, the amplitudes are very small. In \cite{jiao2018physical}, for each side (Alice or Bob), the virtual AoAs and AoDs are estimated and selected as the common randomness. The sparsity reflected in the virtual angle domain makes the estimation process of virtual AoAs/AoDs against noise, which enables virtual AoAs/AoDs estimated at Alice and Bob to be highly identical. Consequently, quantized bits at Alice and Bob have a low bit disagreement ratio within low SNR regimes.

\begin{figure}
\centering
\includegraphics[width=0.48\textwidth,height=0.4\textwidth]{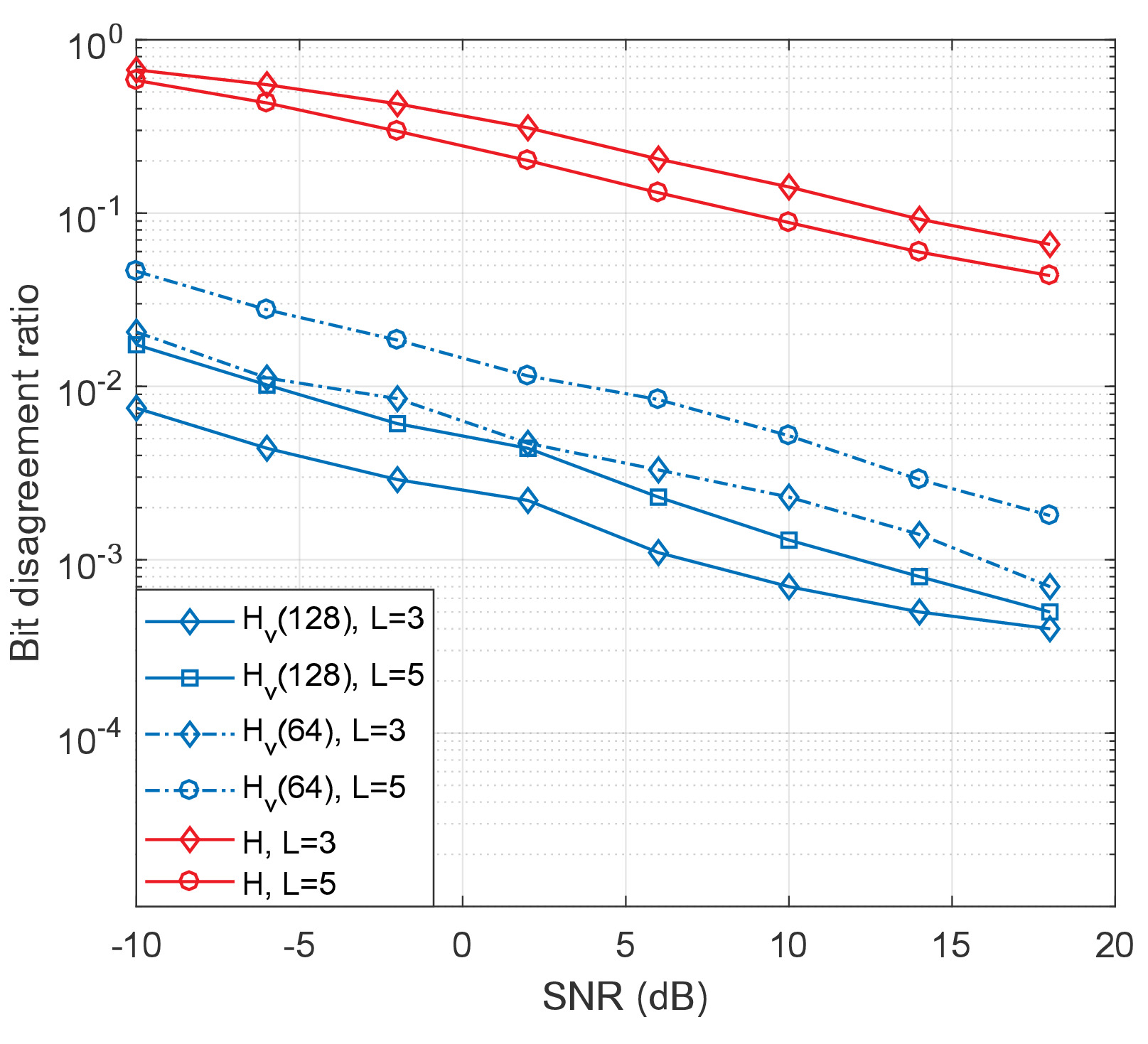}
\caption{Bit disagreement ratio for channel estimate quantization and virtual AoAs and AoDs while $N_t=N_r=128$}
\label{f:KrateHHV}
\end{figure}

Fig. \ref{f:KrateHHV} denotes the performance of the scheme achieving a low bit disagreement ratio.
Using the quantization method in work \cite{MIMOK}, virtual AoAs and AoDs based key generation scheme achieves a very low bit disagreement ratio under low SNR regimes. For example, we compare bit disagreement ratio under two antenna dimensions: $128\times128$ and $64\times64$, where Alice has 128 or 64 antenna elements and Bob has 128 or 64 antenna elements, respectively. The blue curves represent the bit disagreement ratio of the scheme in \cite{jiao2018physical} under different antenna dimensions and multi-path numbers. The red curves represent bit disagreement ratio of existing scheme with the antenna dimensions $128 \times 128$. $L$ represents the number of multi-paths. The numerical result shows that the scheme \cite{jiao2018physical} achieves much lower bit disagreement ratio than the existing scheme \cite{MIMOK}. For instance, when we consider the case where Alice and Bob have 128 antenna elements with $L=3$, the bit disagreement ratio is under $10^{-2}$ even when the SNR is -10dB, which outperforms the existing schemes.


\subsection{Reducing Temporal Correlation under a High Probing Rate Using Multi-resolution Beams}

\label{BeamResolution}

Even though a high channel probing rate provides more samples in a period of time, the key generation rate is not increased linearly to the probing rate. Within the coherence time, a high probing frequency leads to the temporal redundancy along with a large number of repeated bits.
In order to reduce the temporal correlation among samples, the random initial phase can be set to achieve multiple probes in one coherence time. 
However, it cannot be against noise and thus cannot reduce the temporal correlation under a low SNR value.

Hybrid precoding is popular in mmWave massive MIMO channel estimation, where it not only enables the codebook design for mmWave massive MIMO systems but also facilitates beam width adjustment in an easy way.
The beams with adjustable width may focus on different multi-path and angles, which can reduce the correlation among the samples of the channel gain. Thus, we propose a scheme utilizing the multi-resolution beams in the channel probing stage to reduce the temporal correlation as illustrated in Fig. \ref{Hybridprecoding}.
In our simulation, the hybrid precoding scheme is implemented using the hierarchical codebook, while the system operates at 28GHz where Alice has 64 antennas and Bob has 32 antennas and a uniformly quantized phase shifter is considered.
In Fig. \ref{Hybridprecoding}, Alice selects beams with different resolutions or angles to send probing sequences, while Bob uses a fixed beam pattern in the coherence time.
The beams are carefully selected from the hierarchical codebook to maintain the channel gain within a certain range, which enables this scheme to work under the low SNR regime.

\begin{figure}
\centering
\includegraphics[width=0.48\textwidth,height=0.4\textwidth]{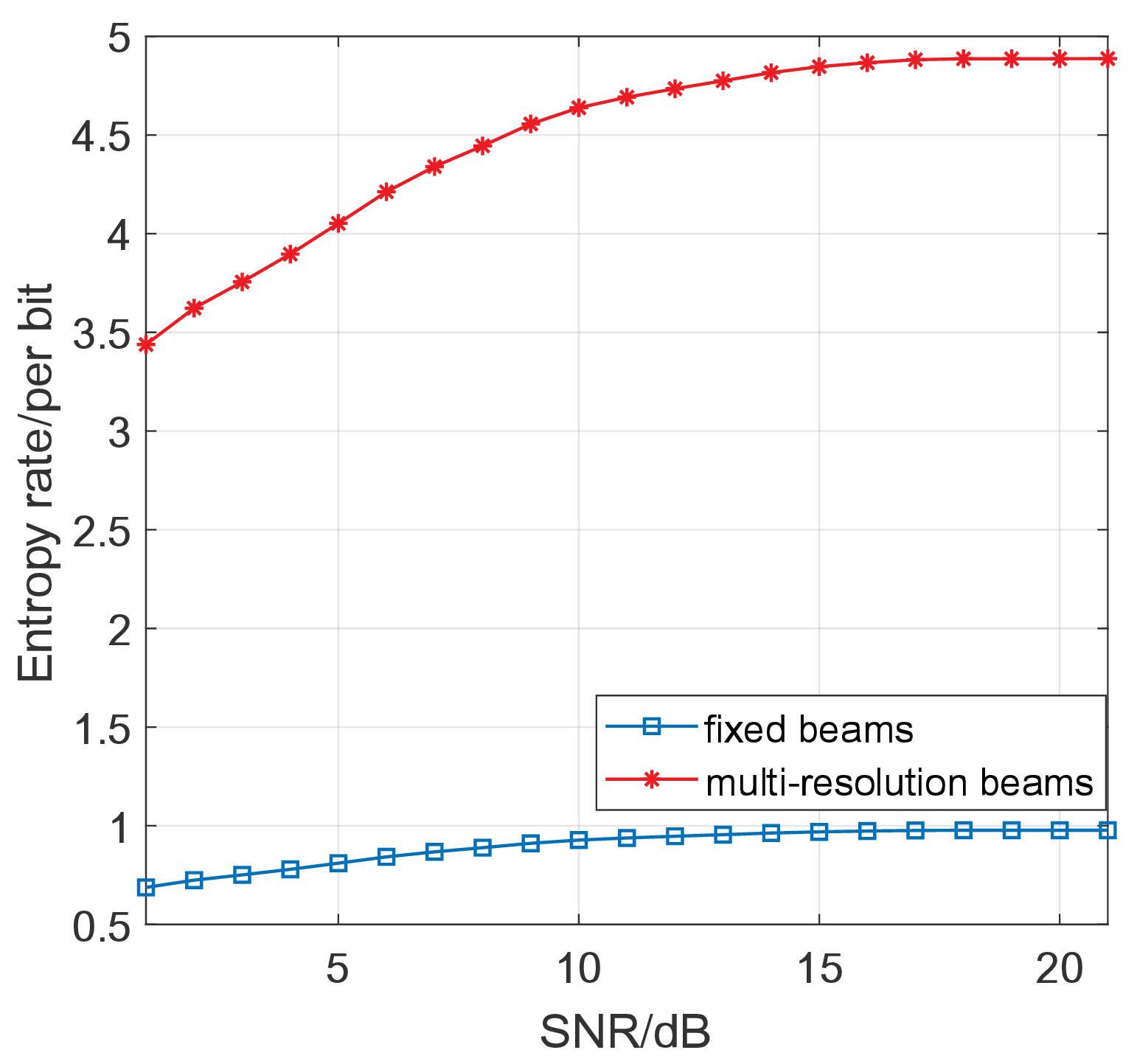}
\caption{Key entropy rate}
\label{f:Temporal}
\end{figure}

Fig. \ref{f:Temporal} illustrates the key entropy rate.
In the coherence time, if we use 5 beams with different resolutions and angles (P=5).
From the figure, we can observe that, the key entropy rate approaches to 1 as the increasing of SNR for the fixed beam. However, our scheme has the key entropy rate approximately 5 times of the fixed beams and thus can reduce the temporal correlation efficiently.
The channel probing with multi-resolution beams can reduce the temporal correlation among samples, so as to further improve the key generation rate with the carefully selected beams combinations.
Besides, the issue that how to maintain the channel gain in a certain range is critical because a big fluctuation of channel gain can affect the key generation rate.

\section{Future Topics}
\label{Future}
In this section, we discuss the potential research directions and future trends of physical-layer key generation for 5G and beyond, including the mobility effect on key generation, multi-user massive MIMO and backscatter communication.

\subsection{The Effect of Mobility on Key Generation}
In mmWave communication, the high path loss and limited scattering effects lead to a high blockage rate.
In such case, the mobility patterns of users may have an effect on physical layer key generation.
How to maintain a stable key generation rate under this case is an open problem.
Besides, beam tracking is developed in mmWave communication to maintain the communication link.
Existing key generation scheme needs to consider the effect of beam tracking.

\subsection{Multi-user Massive MIMO}
To have an efficient system performance, each base station is expected to serve a number of mobile stations, simultaneously.
Thus, hybrid precoding is popular to multiplex different data streams to different users. Digital precoding layer in hybrid precoding structure provides more freedom to reduce the interference among users.
In this case, when it comes to physical layer group key generation problem, hybrid precoding may offer an opportunity to decrease the interference from other users and to maximize the group key generation rate.
However, the problems like how to decrease the channel estimation overhead in the group key generation and how to optimally tune channel probing rate and power need further investigations.

\subsection{Backscatter Communication}
Backscatter communication is an important enabling technology for Internet of Things (IoT) because of its ultra-low power consumption and lower manufacturing cost. Based on this new technique, IoT devices can transmit data by reflecting and modulating incident radio frequency (RF) signals without any power-thirsty RF transmitter. 
In addition, the newest ambient backscatter communications do not rely on any dedicated RF base station, but utilize ambient RF signals, such as TV radio, cellular, and Wi-Fi to communicate with other devices.
However, in such scenarios, the devices cannot directly measure the channel characteristics among themselves as shared randomness to generate secret keys, which makes the conventional physical layer key generation approaches inapplicable. In massive MIMO mmWave backscatter systems, the special channel model, such as round-trip multi-path channel and dyadic backscatter channel, will bring new challenges and opportunities to exact randomness for shared secret key generation. Thus, a new physical key generation schemes in backscatter communications should be considered to obtain a secret key with low computation and communication overhead.

\section{Conclusion}
\label{Conclusion}
In this paper, we reviewed the existing physical layer key generation schemes, and discussed the limitations and opportunities in 5G and beyond.
Through three case studies, including combating co-located eavesdroppers, achieving a low bit disagreement ratio within low SNR regimes, and reducing temporal correlation under high probing rates, we demonstrated the benefits offered by 5G communication technologies for physical layer key generation. We also discussed the future topics and potential research trends related to physical-layer key generation in 5G and beyond.

\bibliography{IEEEabrv,RefAll}
\afterpage{

Long Jiao (ljiao@gmu.edu) received B.Sc. degree in Information Security from Xidian University (XDU), Xian, China, in 2016. He has been with George Mason University, Fairfax, VA, USA, since 2016, where he is currently a Ph.D. student. His current fields of interest include 5G Physical Layer Security, mmWave communication, mmWave HetNets and Deep Learning.
\vspace{3mm}

Ning Wang (nwang5@gmu.edu) is currently a postdoctoral scholar in Electrical and Computer Engineering Department at George Mason University Fairfax, VA, USA. He was an Engineer in Huaxin Post and Telecommunications Consulting Design Co., Ltd., Hangzhou, Zhejiang, Chain, from 2012 to 2013. He received the PhD degree in Information and Communication Engineering from Beijing University of Post and Telecommunication, Beijing, China, in 2017. His current research interests are in physical layer security, machine learning, device identification and RF fingerprinting.
\vspace{3mm}

Pu Wang (pwang20@gmu.edu) received the B.S. degree in Telecommunications Engineering from Xidian University, Xi'an, China in 2014. Currently, He is working toward his Ph.D. in Cyber Engineering at the Xidian University, Xi'an, China. His research interests are in trust management in Internet of Things and cloud computing, anonymous authentication, backscatter communication, wireless information and power transfer, and physical layer security.
\vspace{3mm}

Amir Alipour-Fanid (aalipour@gmu.edu) received the B.S. degree in electrical engineering-power from the Islamic Azad University of Ardabil, Ardabil, Iran, in 2005, and the M.S. degree in electrical engineering-communication from the University of Tabriz, Tabriz, Iran, in 2008. He is currently pursuing the Ph.D. degree with the Electrical and Computer Engineering Department, George Mason University, Fairfax, VA, USA. His research interests include machine learning applications in security and privacy of Cyber-Physical Systems (CPS), Internet of Things (IoT), vehicle-to-vehicle communication, 5G and wireless networks.
\vspace{3mm}

Jie Tang (jtang20@gmu.edu) received the Ph.D. degree from the University of Electronic Science and Technology of China. He is currently a Visiting Scholar at George Mason University, Fairfax, VA, USA. His main interests lie in fundamental mathematical optimization algorithms in wireless communication and networks.
\vspace{3mm}

Kai Zeng (kzeng2@gmu.edu) received the Ph.D. degree in electrical and computer engineering from the Worcester Polytechnic Institute (WPI) in 2008. He was a Post-Doctoral Scholar with the Department of Computer Science, University of California at Davis (UCD) from 2008 to 2011. He was with the Department of Computer and Information Science, University of Michigan-Dearborn as an Assistant Professor from 2011 to 2014. He is currently an Associate Professor with the Department of Electrical and Computer Engineering, Cyber Security Engineering, and the Department of Computer Science at George Mason University. His current research interests are in cyber-physical system security and privacy, 5G physical layer security, network forensics, and spectrum sharing networks. He was a recipient of the U.S. National Science Foundation Faculty Early Career Development (CAREER) Award in 2012. He received the Excellence in Postdoctoral Research Award from UCD in 2011 and the Sigma Xi Outstanding Ph.D. Dissertation Award from WPI in 2008. He is an Editor of the IEEE Transactions on Information Forensics and Security, IEEE Transactions on Wireless Communications, and IEEE Transactions on Cognitive Communications and Networking.

}

\end{document}